\documentclass[12pt]{article}
\usepackage{latexsym}
\usepackage{amsmath}               
\usepackage{amsfonts}              
\usepackage{amsthm}
\usepackage{amssymb}
\usepackage{setspace}
\usepackage{graphicx}
\usepackage{slashbox}
\usepackage{caption}
\usepackage{longtable}
\makeatletter \oddsidemargin .01in \evensidemargin -.1in \textwidth
16cm \topmargin -0.8cm \textheight 22.05cm
\usepackage{algorithm}
\usepackage{algorithmic}

\newcommand{\bea}{\begin{eqnarray}}
\newcommand{\eea}{\end{eqnarray}}

\begin{document}


\title{\bf\large A Study On ID-based Authentication Schemes for Telecare Medical Information System}


\author{\small Dheerendra Mishra\thanks{E-mail:~{dheerendra@maths.iitkgp.ernet.in} }\\
\small Department of Mathematics\\
\small Indian Institute of Technology Kharagpur\\
\small  Kharagpur 721302, India\\}

\date{}
 \maketitle

\begin{abstract}

The smart card based authentication schemes are designed and developed to ensure secure and authorized communication between remote user and the server. In recent times, many smart card based authentication schemes for the telecare medical information systems (TMIS) have been presented. In this article, we briefly discuss some of the recently published smart card based authentication schemes for TMIS and try to show why efficient login and password change phases are required. In other word, the study demonstrates how inefficient password change phase leads to denial of server attack and how inefficient login phase increase the communication and computation overhead and decrease the performance of the system.


\end{abstract}
\textbf{keywords:} {Telecare medical information system; Smart card; Password based authentication; Login phase; Password change phase.}

\section{Introduction}{\label{intro}}

 User-friendly, omnipresence and low cost of internet technology, facilitates flexible online services in E medicine in which a user can access the remote services at any instant from anywhere. One of service in  C-medicine is telemedicine, which has been brought to patients' home through the telecare medical information system (TMIS). In TMIS, medical server maintains the electronic medical records (EMRs) of registered users and facilitates various services to the users, health educators, physicians, hospitals, care-givers, public health organizations and home-care service providers.

 In Telecare system a user access the server via public channel while the adversary may have full control over the public channel and so he can eavesdropped, intercept, record, modify, delete, and replay the message broadcasting via public channel~\cite{xu2009improved}. This cause data privacy and security threat in the deployment of TMIS.

The smart card based protocols enable the health-care delivery services door to door by employing information and communication technologies~\cite{kumari2012cryptanalysis,lee2013secure,martin2004synergy}. TMIS has the capability to reduce the social and medical expenses. It also improves the quality of  medical services quality and efficiency. It is regarded as an efficient, economical and time saving alternative compare to traditional clinical service~\cite{lin2012development,kukafka2007redesigning,pizziferri2005primary}. However, the privacy and security issues are rising in TMIS that related to the security of transmitted data and the rights of patients to understand and control the use of their presentational health records. To ensure security and privacy during information transmission via public channel, the smart card based anonymous remote user authentication schemes are generally adopted. In general, smart card based authentication faces various attacks, a few are, stolen smart card attack, password guessing attack and insider attack~\cite{boyd2003protocols}.

Recently, a number of password based schemes presented for TMIS~\cite{chen2012efficient,debiao2012more,lee2013secure,pu2012strong,wei2012improved,wu2012secure,xie2013robust,zhu2012efficient,xu2014secure}.
In 2010, Wu et al.~\cite{wu2012secure} presented an efficient authentication scheme for TMIS. Their scheme is better than the previously proposed schemes for low computing devices by adding pre-computing phase.  In pre-computing phase, user performs exponential operation and then stores the calculated values into the storage device, such that a user can extract these value from the device whenever required. However, Debiao et al.~\cite{debiao2012more} demonstrated that Wu et al.'s scheme fails to resist the impersonation attack to the insider attack. They also introduced an enhanced scheme and claimed that their proposed scheme eliminates the drawbacks of Wu et al.'s scheme and performs better than Wu et al.'s scheme.  They also claimed that their scheme is more appropriate for low power mobile devices for TMIS. In 2012, Wei et al.~\cite{wei2012improved} identified that both Wu et al.'s and Debiao et al.'s schemes are inefficient to meet two-factor authentication. In general, an efficient smart card based password authentication scheme should support two-factor authentication. They also presented an improved smart card based authentication scheme for TMIS and claimed that the improved scheme is efficient and achieved two-factor authentication. However,  Zhu~\cite{zhu2012efficient} demonstrated that Wei et al.'s scheme is vulnerable to off-line password guessing attack using stolen smart card.  He also presented an improved scheme for TMIS and claimed that his scheme could overcome the weaknesses of Wei et al.'s scheme. Recently, Lee and Liu~\cite{lee2013secure} showed that Zhu's scheme does not resist parallel session attacks and introduced an improved scheme. They claimed that their scheme provides secure and efficient solution for telecare medical information system.

 In recent time Chen et al.~\cite{chen2012efficient} also presented an efficient and secure dynamic ID-based authentication scheme for TMIS. In 2013, Lin~\cite{lin2013security} demonstrated that in Chen et al's scheme, user identity is compromised under the dictionary attack and  password  can be derived with the stolen smart card. Cao and Zhai~\cite{cao2013improved} demonstrated that Chen et al's scheme is vulnerable to an off-line identity guessing attack, an off-line password guessing attack, and an undetectable on-line password guessing attack when user's smart card is stolen. They also presented an improved scheme for TIMS. Xie et al.~\cite{xie2013robust} also pointed out the weaknesses of Chen et al.'s scheme and presented an improved scheme for TIMS. Recently, Xu et al.~\cite{xu2014secure} introduced ECC based authentication scheme for Telecare Medicine Information Systems. Most of the existing schemes for TMIS have lack of input verifying condition which makes these schemes inefficient. In this article, we  show that recently presented authentication schemes~\cite{lee2013secure,lin2013security,wei2012improved,xie2013robust,zhu2012efficient,xu2014secure} fail to provide  efficient login and password change phase.

The rest of the paper is organized as follows: Section \ref{rw} presents the brief review of Wei et al.'s scheme.  Section \ref{cw} points out the flaws in login and password change phase of Wei et al.'s scheme. Section \ref{rz} \& \ref{cz} briefly discuss the Zhu's scheme and show their weaknesses. Section \ref{rl} presents the brief review of Lee Liu's scheme and the analysis is presented in Section \ref{cl}.  Section \ref{lin-review} presents the brief review of Lin's scheme.  Section \ref{lin-crypt} points out the drawbacks of inefficient login and password change phase of Lin's scheme. Section \ref{rc} presents the brief review of Cao and Zhai's scheme.  Section \ref{cc} points out the weakness of Cao and Zhai's scheme. A brief review of the Xie et al.'s scheme is given in Section \ref{rx} and scheme loop holes are presented in Section \ref{cx}. Section \ref{xu-review} presents the brief review of Xu et al.'s scheme.  Section \ref{lin-crypt} points out the drawbacks of inefficient login and password change phase of Xu et al.'s scheme. The attribute comparison  is presented in Section \ref{comparision}. Finally, the conclusion is drawn in Section \ref{conclusion}.

%
%
%
%
%


\section{Notations}

\begin{table}[H]
  \centering
 \caption{Meaning of symbols that are used in this article}\label{t1}

\begin{tabular}{l|l}
\hline
Notation & Descryption\\
\hline
$U$ & User/paitient\\
$S$ & A trustworthy medical server\\
$p, q$ & Two large primes\\
$N$ & Product of $p$ and $q$\\
$Z_N$ & Ring of integers modulo $N$\\
$e$ & Public key of $S$\\
$d$  & Master secret key of $S$\\
$E$ & Attacker/adversary\\
$SC$ & Smart Card\\
$ID_U$ & Unique identity of $U$\\
$PW_U$  & Unique password of $U$\\
$E_X$ & Symmetric key encryption using key $X$\\
$x$  & Secret value (master key) of $S$\\
$h(\cdot)$ & A collision resistant one-way hash function\\
$\oplus$ & XOR\\
$||$ & String concatenation operation\\
\hline
\end{tabular}

\end{table}

\section{Review of Wei et al.'s Scheme}\label{rw}
In Wei et al.'s scheme~\cite{wei2012improved}, the remote telecare server is
responsible for generating global parameters, computing
users secret information and providing smart cards to the
new users. There are four phases in their scheme:\\
1.) Initial Phase\\
2.) Registration phase\\
3.) Authentication phae\\
4.) Password change phase


\subsection{Initial Phase}

Initially, the telecare server defines its parameters. It chooses two large primes $p$ and $q$ such that $p= 2q+1$. It also selects a generator $g$ of $Z^{*}
_q$ and the system master key $x \in  Z^{*}_q$. It defines a one-way hash function $h(\cdot) :\{0, 1\}^{*}\rightarrow  Z^{*}_q$.

\subsection{Registration Phase}
To access the telecare services, a patient registers with the server and achieves the personalized smart card as follows:

\begin{description}

\item [\bf Step 1.]
$U$ generates a random number $N$, selects the identity $ID_U$ and password $PW_U$ and computes $W = h(PW_U\parallel N)$. Then $U$ sends $ID_U$ and $W$ to  $S$.

\item[\bf Step 2.] Upon receiving $ID_U$ with $W$, $S$ calculates $B = h(ID_U^x) + h(W) \pmod p$. Then, $S$ embeds $h(\cdot), g, p, ID_U$ and $B$
into the smart card and provides it to $U$.

    \item [\bf Step 3.] Upon receiving the smart card, $U$ stores $N$ into the smart card.

\end{description}

\subsection{Authentication  Phase}
When a legal user wishes to login into TMIS, the following steps are executed.


\begin{description}

\item [\bf Step 1. ] $U$ inserts the smart card into a card reader and inputs $PW_U$. Then, $SC$ generates a random number $r_U$, computes  $A = g^{r_U} \pmod p$, $W = h(PW_U\parallel N), B^{'}=(B - h(W))$*$A$ and $h_1= h(A\parallel B^{'}\parallel ID_U)$. Then, $U$ sends the message $M_1= \{ID_U, B', h_1\}$ to the server $S$.

\item [\bf Step 2. ] Upon receiving the message, $S$ verifies the validity of $ID_U$. If $ID_U$ is invalid, $S$ terminates the session. Otherwise, $S$ calculates $A^{'} = \frac{B^{'}}{h(ID^x)}\pmod p$ and verifies $h_1 =?~h(A\parallel B' \parallel ID_U)$. If verification does not hold, $S$ stops the session. Otherwise, it generates a random number $r_S$ and calculates $sk= h(ID_U\parallel A^{'}\parallel B^{'} \parallel r_S)$ and $h_2= h(sk\parallel r_S)$. Then, $S$ sends the message $M_2= \{h_2, r_S\}$ to $U$.

\item[\bf Step 3.] Upon receiving the message $M_2$, $SC$ calculates $sk' = h(ID_U\parallel A\parallel B' \parallel r_S)$ and verifies $h_2 =?~h(sk'\parallel r_S)$. If verification fails, it  stops the session. Otherwise, $SC$ computes $h_3 = h(ID_U\parallel sk^{'})$ and sends $M_3= \{h_3\}$ to $S$.

\item[\bf Step 4.] Upon receiving $M_3$, $S$ verifies $h_3 =?~h(ID_U\parallel sk)$. If verification fails, $S$ stops the session. Otherwise, $U$ is authenticated.
\end{description}

\subsection{Password Change Phase}

Any legal user $U$ can change the password by using the following steps.
\begin{description}

\item [\bf Step 1. ] $U$ inserts his smart card into a card reader and inputs the old password $PW_U$ and the new password $PW_{new}$.
\item [\bf Step 2. ] The smart card computes $W = h(PW_U\parallel N), W_{new} = h(PW_{new}\parallel N)$ and $B_{new} = B - h(W) + h(W_{new}) \pmod p$.
\item [\bf Step 3. ] Finally , the smart card replaces $B$ with $B_{new}$.
\end{description}
%


\section{Analysis of Wei et al.'s Scheme}\label{cw}

 In 2012, Zhu's~\cite{zhu2012efficient} demonstrated that Wei et al.'s scheme does not resist offline password guessing attack. We revisits the Wei et al.'s scheme and find some more flaws. In this section, we show that Wei et al.'s scheme does not provide efficient login and password change phase.

\subsection{Inefficient password change phase}
An efficient smart card based remote user authentication scheme should ensure efficient password change phase where incorrect input can be quickly detected and user can change his password without server assistance. However, in Wei et al.'s scheme, if a user inputs wrong old password, the smart card does not verify the correctness of password and executes the password change phase. Although a user may enter a wrong password, as human may sometime forgot the password or commit some mistake. This may cause the denial of service where a user will no longer communicate with server using the same smart card. If user inputs wrong password $PW_U^*$ instead of $PW_U$, then the smart card executes the password change phase as follows:

  \begin{itemize}
    \item  Compute $W^* = h(PW_U^*||N)$ and then
    \begin{eqnarray*}
    B_1 &= &B - h(W^*)\\
     &=  &h(ID_U^x) + h(h(PW_U||N)) - h(h(PW_U^*||N)) \neq h(ID_U^x),
\end{eqnarray*}
as $PW_U^*\neq PW_U$.
    \item Compute
    \begin{eqnarray*}
    B_{new}& = &B_1 + h(h(PW_{new}||N))\\
     &= &h(ID_U^x) + h(h(PW_U||N)) - h(h(PW_U^*||N)) + h(h(PW_{new}||N)).
\end{eqnarray*}
    \item  Replace $B$ with $B_{new}$.

  \end{itemize}

The smart card inability of wrong password detection, may cause \textbf{denial of service} as follows:
 \begin{itemize}

 \item $U$ inputs the current password $PW_{new}$ into smart card. Then, $SC$ generates a random number $r_U$, computes  $A = g^{r_E} \pmod p$, $W_{new} = h(PW_{new}||N)$, then achieves $B'_{new}$ as follows:

     \begin{eqnarray*}
       B'_{new} &=& (B_{new} - h(W_{new}||N))*A\\
                 &=& (h(ID_U^x) + h(h(PW_U||N)) - h(h(PW_U^*||N)))*A.
     \end{eqnarray*}

 \item SC also computes $h_1= h(A\parallel B'_{new}\parallel ID_U)$. Then, it sends the message $M_1= \{ID_U, B'_{new}, h_1\}$ to the server $S$.

\item Upon receiving the message, $S$ verifies the validity of $ID_U$.  The verification holds, as message includes user's original identity. Then, $S$ calculates $A^{'} = \frac{B'_{new}}{h(ID^x)}\pmod p \neq A,$ as $B'_{new} = (h(ID_U^x) + h(h(PW_U||N)) - h(h(PW_U^*||N)))*A$ instead of $(h(ID_U^x))*A$ where $PW_U\neq PW_U^*$.

   \item $S$ verifies $h_1 =?~h(A'\parallel B'_{new}\parallel ID_U)$. The verification does not hold, as $A'\neq A$. Then, $S$ terminates the session, as message authentication fails.

\end{itemize}
In this case, user can never login to the server using the same smart card, which causes \textbf{denial of service attack}.

\subsection{Inefficient login phase:}
An efficient smart card based remote user authentication scheme should be able to identify incorrect login request so that extra communication and computation cost should not occur due to incorrect login. However, Wei et al.'s scheme does not verify the correctness of input and executes the session. If  a user inputs incorrect input, then the following cases are possible:

\noindent{\bf Case 1.}
 If user inputs  wrong password $PW_U^*$ instead of $PW_U$:

\begin{itemize}
 \item $SC$ generates a random number $r_U$, computes  $A = g^{r_U} \pmod p$, $W^* = h(PW_U^*\parallel N)$ and $B'^* = (B - h(W^*))*A$ as


    \begin{eqnarray*}
      (B - h(W^*))*A &=& (h(ID_U^x) + h(h(PW_U||N)) - h(h(PW_U^*||N)))*A \\
       & \neq& h(ID_U^x)*A, as  PW_U^* \neq PW_U
    \end{eqnarray*}

Then, it sends the message $M_1= \{ID_U, B'^*, h_1^*\}$ to $S$, where $h_1^* = h(A\parallel B'^* \parallel ID_U)$.

\item  Upon receiving the message, $S$ verifies the validity of $ID_U$.  The verification holds, as message includes user's original identity. Then, $S$ calculates $A' = \frac{B'^*}{h(ID^x)}\pmod p\neq A$, as $B'^* = (h(ID_U^x) + h(h(PW_U||N)) - h(h(PW_U^*||N)))*A$ instead of $h(ID_U^x)*A$.

      \item When $S$ verifies $h_1^* =?~h(A'\parallel B'^* \parallel ID_U)$. The verification does not hold, since $A\neq A'$.  Then $S$ terminates the session as message authentication does not hold.

\end{itemize}



\noindent{\bf Case 2.} An attacker can also intercept and modify the login message:

\begin{itemize}
  \item An attacker intercepts $U$'s login message  $M_1= \{ID_U, B', h_1\}$, then selects a number $r_E$ and computes $B'^* = B'r_E$. Finally, $E$ replace $M_1$ with $M_1^*$, where $M_1^* = \{ID_U, B'^*, h_1\}$.

  \item On receiving the message $M_1^*$, when $S$ verifies $h_1 =?~h(A'\parallel B'^* \parallel ID_U)$. The verification fails because $B'^* = B'*r_E$ instead of $B'$. As a result, $S$ terminates the session.
\end{itemize}

In both the cases, wrong password input and message impersonation attack, session terminates. And, user may not identify the correct reason of session failure.
%


\section{Review of Zhu's Scheme}\label{rz}

In 2012, Zhu~\cite{zhu2012efficient}  proposed an improvement of Wei el al.'s~\cite{wei2012improved}. Zhu's scheme comprises four phases similar to the Wei et al.'s scheme. The brief description of Zhu's scheme is as follows:

\subsection{Initial Phase}

The remote telecare server $S$ executes the following steps to initiate the session:\\
\begin{description}

\item [\bf Step 1.] $S$ generates two large primes $p$ and $q$ and computes $n=pq$.

\item[\bf Step 2.] $S$ chooses a prime number $e$ and an integer $d$ as system public and private keys such that $ed\equiv 1 \pmod {(p-1)(q-1)}$ and a one-way
hash function $h(\cdot): \{0\}^*\rightarrow Z^*_q
$, where $e$ and $d$  should be provided to the server in a safe
way.

\end{description}
\subsection{Registration Phase}

Registration phase carries out the following steps:\\
\begin{description}

\item [\bf Step 1.] $U$ generates a random number $N$, selects his identity
$ID_U$ and password $PW_U$ and computes $W = h(PW_U\parallel N)$. Then $U$ sends $ID_U$ and $W$ to the
telecare server $S$.

\item[\bf Step 2.] Upon receiving the registration request, $S$ computes $B = h(ID_U \oplus d)\oplus h(W)$. Then, $S$ stores $n, e, ID_U$ and $B$
into $U$'s smart card and issue the smart card to $U$.

    \item [\bf Step 3.] After receiving the smart card, $U$ stores $N$ into the smart card. Finally, the smart card includes the following security parameters $\{n, e, ID_U, B, N\}$.

\end{description}

\subsection{Authentication  Phase}
User and server can verify the authenticity of each other by adopting the following procedure:

\begin{description}

\item [\bf Step 1. ] $U$ inserts his smart card into a card reader and inputs $PW_U$. Then $SC$ generates a random number $r_U$, computes $W = h(PW_U\parallel N), B' = B\oplus h(W), h_1= h(B'\parallel r_U)$ and $X = {(h_1\parallel r_U)}^e\pmod n$. Finally, $U$ sends the message
$M_1= \{ID_U, X\}$ to the server $S$.

\item [\bf Step 2. ] Upon receiving $M_1$, $S$ checks the validity of $ID_U$. If $ID_U$ is not valid, $S$ terminates the session. Otherwise, $S$ achieves $h'_1\parallel r_U = X^d\pmod n$ and verifies $h_1^{'}= h(h(ID_U \oplus d)\parallel r_U')$. If it does not hold, $S$ terminates the session. Otherwise, runs {\em Step 3}.

    \item[\bf Step 3. ] $S$ generates a random number $r_U$, calculates $h_2= h(ID_U, r_U', r_S)$ and transmits $M_2= \{h_2, r_S\}$ to $S$.

\item[\bf Step 4.] Upon receiving the message $M_2$, $SC$ verifies $h_2 =?~h(ID_U, r_U, r_S)$. If verification does not hold, the smart card terminates the session. Otherwise, it calculates $h_3 = h(ID_U, r_S, r_U)$ and sends the message $M_3= \{h_3\}$ to $S$.

\item[\bf Step 4.] Upon receiving $M_3$, $S$ verifies $h_3 =?~h(ID_U, r_S, r_U')$. If verification fails, $S$ terminates the session. Otherwise, $U$ is authenticated.
\end{description}
\subsection{Password Change Phase}

Any legal user with valid smart card can change the password as follows:
\begin{description}

\item [\bf Step 1. ] $U$ inserts his smart card into a card reader and inputs the old password $PW_U$ and the new password $PW_{new}$.
\item [\bf Step 2. ] $SC$ computes $W = h(PW_U\parallel N), W_{new} = h(PW_{new}\parallel N)$ and $B_{new} = B\oplus h(W) \oplus h(W_{new})$. Finally, it replaces $B$ with $B_{new}$.

\end{description}

\section{Analysis of Zhu's Scheme}\label{cz}
In 2013, Lee and Liu~\cite{lee2013secure} demonstrated parallel session attack on Zhu's Scheme~\cite{zhu2012efficient}. In this section, we will show that Zhu's scheme fails to provide efficient login and password change phase.



\subsection{Inefficient password change phase}
  User can change his password independently without server's assistance. However, the smart card does not verify the user's password. If user inputs wrong password $PW^*$ instead of $PW$, then the smart card executes the password change phase as follows:

  \begin{itemize}
    \item  Compute $W^* = h(PW_U^* \parallel N)$ and then $B^*$ as:

     \begin{eqnarray*}
       B^* &=& B\oplus h(W^*)\oplus h(PW_{new}\parallel N) \\
           &=& h(ID_U\oplus d)\oplus h(PW_U\parallel N)\oplus h(PW_U^*\parallel N)\oplus h(PW_{new}\parallel N)\\
           &\neq&  h(ID_U\oplus d)\oplus h(PW_{new}\parallel N),~ as~ PW_U\neq PW_U^*.
     \end{eqnarray*}

    \item  Replace $B$ with $B^*$.

  \end{itemize}
The password change phase succeeds, in case of wrong password input. This may cause {\em denial of service}, which is clear from the following facts:

 \begin{itemize}

 \item $U$ inputs $PW_{new}$ into smart card. Then, smart card generates a random number $r_U$ and computes $W = h(PW_{new}\parallel N)$, then $H^*$ as follows:

   \begin{eqnarray*}
     H^* &=& B^*\oplus h(W_{new}) \\
         &=& h(ID_U\oplus d)\oplus h(PW_U\parallel N)\oplus h(PW_U^*\parallel N)\oplus h(PW_{new}\parallel N)\oplus h(PW_{new}\parallel N) \\
         &=& h(ID_U\oplus d)\oplus h(PW_U\parallel N)\oplus h(PW_U^*\parallel N)\\
         &\neq& h(ID_U\oplus d)~=~H,~ as~ PW_U \neq PW_U^*
   \end{eqnarray*}

 Then, it achieves $h_1^* = h(H^*||r_U), X^* = (h_1^*||r_U)^{e}\pmod n$, and sends $M_1^* = \{ID_U, X^*\}$ to $S$.

\item  Upon receiving $M_1^*$, $S$ achieves $h_1^{*}||r_U = {X^*}^{d}\pmod n$ and then verifies the validity of $ID_U$. The verification holds. However, when it computes $H = h(ID_U\oplus d))$ and checks $h_1^* = h(H||r_U)$. The verification does not hold, as $H\neq H^*$.
\item $S$ terminates the session as message authentication fails.

\end{itemize}
This is denial of service condition.

\subsection{Inefficient login phase:}
A user may input wrong password in login phase, as a human may commit mistake. However, in Zhu's scheme, the smart card executes the session in case of wrong password inputs.

\noindent{\bf Case 1.}
 If user inputs wrong password $PW_U^*$, then smart card executes the login phase without verifying the correctness of password as follows:

\begin{itemize}
 \item The smart card generates a random number $r_U$ and computes $W = h(PW_U^*\parallel N)$, then $H^*$ as follows:

   \begin{eqnarray*}
     H^* &=& B\oplus W^* \\
         &=& h(ID_U\oplus d)\oplus h(PW_U\parallel N)\oplus h(PW_U^*\parallel N)\\
         &\neq& h(ID_U\oplus d)~=~H,~ as~PW_U \neq PW_U^*
   \end{eqnarray*}

 Then, it achieves $h_1^* = h(H^*||r_U), X^* = (h_1^*||r_U)^{e}\pmod n$, and sends $M_1^* = \{ID_U, X^*\}$ to $S$.

\item  Upon receiving $M_1^*$, $S$ achieves $h_1^{*}||r_U = {X^*}^{d}\pmod n$ and then verifies the validity of $ID_U$. The verification holds. However, when it computes $H = h(ID_U\oplus d))$ and checks $h_1^* = h(H||r_U)$. The verification does not hold, as $H\neq H^*$.

\item $S$ terminates the session as message authentication fails.

\end{itemize}


\section{Review of Lee and Liu's Scheme}\label{rl}

Lee and Liu~\cite{lee2013secure} presented the enhanced authentication scheme based on the scheme of Zhu's scheme~\cite{zhu2012efficient} to eliminates the weaknesses of Zhu's scheme. This scheme comprises of initial phase, registration, authentication and key agreement and password change phase. Since the initialization phase is same as Zhu's scheme. So, we only briefly discuss registration, authentication and key agreement and password change phase, which is as follows:


\subsection{Initial Phase}

This phase is similar to the Zhu's scheme~\cite{zhu2012efficient}.

\subsection{Registration Phase}

This phase is also similar to the Zhu's scheme, the only difference is that the smart card stores one more parameter  a serial number $SN_U = 0$. Finally, the smart card stores the values $\{n, e, ID_U, B, N, SN_U\}$.

\subsection{Authentication  Phase}
To verify the authenticity of each other the server and user perform the following steps:

\begin{description}

\item [\bf Step 1. ] $U$ inserts the smart card into a card reader and inputs his password $PW_U$. Then, SC generates a random number $r_U$, computes $W = h(PW_U||N), SN_U = SN_{U+1}, H = B\oplus W, h_1= h(H||r_U||SN_U), X= (ID_U||h_1||r_U||SN_U)^{e}\pmod n$, and sends $M_1= \{X\}$ to $S$.

\item [\bf Step 2. ] Upon receiving $M_1$, $S$ achieves $ID_U||h_1^{'}||r_U'||SN_U^{'} = X^{d}\pmod n$ and then verifies the validity of $ID_U$ and
check $SN_U^{'}> SN$, $h_1^{'}= h(h(ID_U\oplus d)||r_U'||SN_U^{'})$, where $SN$ is initialized as $0$ and stores in $S$ database. If verification does not hold, terminates the session. otherwise, $S$ generates a random number $r_S$ and computes $h_2 = h(ID_U||r_U'||r_S||SN_U^{'})$ and sends $M_2 = <h_2, r_S\oplus r_U'>$ to $U$. $S$ also updates $SN$ with $SN_U^{'}$ and keeps it until this session is completed.

\item[\bf Step 3.] Upon receiving $M_2$, $U$ achieves $r_S' = r_S\oplus r_U'\oplus r_U$  and verifies $h_2 =?~h(ID_U||r_U||r_S'||SN_U)$. If
 verification fails, $U$ stops the session. Otherwise, $U$ calculates $sk = h(ID_U||r_S'||r_U||SN_U)$ and $h_3 = h(SK)$, then sends $M_3 = <h_3>$ to $S$.

\item[\bf Step 4.] Upon receiving $M_3$, $S$ computes $sk^{'}= h(ID_U ||r_S||r_U'||SN_U^{'})$ and verifies $h_3 =?~h(sk^{'})$. If verification fails, $S$ stops the session. Otherwise, $S$ consider $U$ as authorized user  and $sk= sk^{'}$ as the session key.
\end{description}
\subsection{Password Change Phase}

A user $U$ with valid smart card a correct password can change the password as follows:
\begin{description}

\item [\bf Step 1. ] $U$ insert the smart in the card reader with old password $PW_U$ and new password $PW_{new}$.
\item [\bf Step 2. ] The smart card computes $W = h(PW_U\parallel N), W_{new} = h(PW_{new}\parallel N)$ and $B_{new}= B \oplus W \oplus W_{new}$.
\item [\bf Step 3. ]The smart card replaces $B$ with $B_{new}$
.
\end{description}


\section{Analysis of Lee and Liu's Scheme}\label{cl}

\subsection{Inefficient login phase:}
In Lee and Liu's scheme, the smart card executes the session without verifying the correctness of input. If a user inputs wrong password $PW_U^*$, then smart card executes the login phase without verifying the correctness of password as follows:

\begin{itemize}
 \item The smart card generates a random number $r_U$ and computes $W = h(PW_U^*\parallel N)$, then $H^*$ as follows:

   \begin{eqnarray*}
     H^* &=& B\oplus W^* \\
         &=& h(ID_U\oplus d)\oplus h(PW_U\parallel N)\oplus h(PW_U^*\parallel N)\\
         &\neq& h(ID_U\oplus d)~=~H,~ as~PW_U \neq PW_U^*
   \end{eqnarray*}

 Then, it achieves $h_1^* = h(H^*||r_U||SN_U), X^* = (ID_U||h_1^*||r_U||SN_U)^{e}\pmod n$, and sends $M_1^* = \{X^*\}$ to $S$.

\item  Upon receiving $M_1^*$, $S$ achieves $ID_U||h_1^{*}||r_U||SN_U = {X^*}^{d}\pmod n$ and then verifies the validity of $ID_U$ and
check $SN_U^{'}> SN$. The verification holds. However, when it computes $H = h(ID_U\oplus d))$ and checks $h_1^* = h(H||r_U||SN_U)$. The verification does not hold, as $H\neq H^*$.
\item $S$ terminates the session as message authentication fails.

\end{itemize}

If a user will try again and again by inputting the wrong password. However, the increase number trials would flood up the network with fake requests. Additionally, it will also cause the server to indulge in unnecessary calculations causing wastage of time and resources.

\subsection{Inefficient password change phase}
 In Lee and Liu's scheme, the user can change his password independently without server's assistance. However, the smart card does not verifies the user's password password. When a user inputs the wrong password, the Smart card does not verifies the correctness of password and process the password chance request. However, a user may enter a wrong password, as human may sometime forgot the password or commit some mistake. This may cause the denial of service scenario where a user will no longer communicate with server using the same smart card. If user inputs wrong password $PW_U^*$ instead of $PW_U$, then the password change phase works as follows:

  \begin{itemize}
    \item  Compute $W^* = h(PW_U^*\parallel N)$ and then $B^* = B\oplus W^*\oplus h(PW_{new}\parallel N) = h(ID_U\oplus d)\oplus h(PW_U\parallel N)\oplus h(PW_U^*\parallel N)\oplus h(PW_{new}\parallel N)\neq  h(ID_U\oplus d)\oplus h(PW_{new}\parallel N)$ as $PW_U\neq PW_U^*$.

    \item  Replace $B$ with $B^*$.

  \end{itemize}
The password change phase succeeds, in case of wrong password input. This may cause {\em denial of service attack}, which is clear from the following facts:

 \begin{itemize}

 \item When $U$ inputs new password $PW_{new}$ into smart card for login. Then, smart card generates a random number $r_U$ and computes $W_{new} = h(PW_{new}\parallel N)$, then $H^*$ as follows:

   \begin{eqnarray*}
     H^* &=& B^*\oplus W_{new} \\
         &=& h(ID_U\oplus d)\oplus h(PW_U\parallel N)\oplus h(PW_U^*\parallel N)\oplus h(PW_{new}\parallel N)\oplus h(PW_{new}\parallel N) \\
         &=& h(ID_U\oplus d)\oplus h(PW_U\parallel N)\oplus h(PW_U^*\parallel N)\\
         &\neq& h(ID_U\oplus d)~=~H,~ as~ PW_U \neq PW_U^*
   \end{eqnarray*}

 Then, it achieves $h_1^* = h(H^*||r_U||SN_U), X^* = (ID_U||h_1^*||r_U||SN_U)^{e}\pmod n$, and sends $M_1^* = \{X^*\}$ to $S$.

\item  Upon receiving $M_1^*$, $S$ achieves $ID_U||h_1^{*}||r_U||SN_U = {X^*}^{d}\pmod n$ and then verifies the validity of $ID_U$ and
check $SN_U^{'}> SN$. The verification holds. However, when it computes $H = h(ID_U\oplus d))$ and checks $h_1^* = h(H||r_U||SN_U)$. The verification does not hold, as $H\neq H^*$.
\item $S$ terminates the session as message authentication fails.

\end{itemize}
In this case, user can never login to the server. This shows that denial of service attack succeeds if user inputs wrong password in password change phase.


\section{Review of Lin's Dynamic ID-based Authentication Scheme for Telecare Medical Information Systems}\label{lin-review}

In 2013, Lin~\cite{lin2013security}  proposed an improvement of Chen el al.'s dynamic ID-based authentication scheme for TMIS~\cite{chen2012efficient}. Their scheme is motivated by Zhu's scheme~\cite{zhu2012efficient} and based on RSA public-key cryptosystem \cite{rivest1978method}. In their scheme, medical server $S$ chooses two large
primes $p$ and $q$ then computes $N = pq$. $S$ also chooses an integer $e$ which is relatively prime to $(p-1)(q-1)$ and derives $d$ such that $ed \equiv 1$ (mod
$(p-1)(q-1))$. Lin's scheme consists registration, login, verification and password change chases which is similar to Chen et al.'s scheme. The brief description of registration, login, verification and password change phases are given below:

\subsection{Registration Phase}

A user $U$ registers his identity $ID_U$ to the medical server $S$ and gets the medical smart card $SC$ with personalized security parameters as follows:
\begin{description}

\item [\bf Step 1.] User $U$ selects a password and an integer $t\in Z_N$. $U$ computes $W = h(PW_U\oplus t)$ and sends the message $<ID_U, W>$ to $S$ via secure channel.

\item[\bf Step 2.] Upon receiving $U$'s message $<ID_U, W>$, $S$ computes $v = W\oplus h(d\oplus ID_U)$. Then $S$ personalizes $U$'s medical smart card by embedding the parameters $\{N, v, e\}$ and returns the medical smart card to $U$ via secure channel.

    \item [\bf Step 3.] Upon receiving the smart card, $U$ stores $t$ into the smart card, i.e., smart card stores $\{N, v, e, t\}$.

\end{description}

\subsection{Login Phase}
  The user $U$ inserts his smart card $SC$ into the smart card reader, and inputs $ID_U$ and $PW_U$. $SC$ chooses $k\in Z_N$ and computes $W = h(PW_U\oplus t)$, $H = v\oplus W$, $CID = h(H\oplus k)$, $R = h(CID, k, ID_U, T_1)$ and $X = (CID||k||ID_U)^e~\mbox{mod}~N$ where $T_1$ is the current timestamp. Finally, the message $<X, R, T_1>$ is sent to $S$.

\subsection{Verification Phase}
In verification phase, user $U$ and medical server $S$ verify the authenticate each other and compute session key as follows:


\begin{description}

\item [\bf Step 1.] Upon receiving $U$'s message $<X, R, T_1>$, $S$ verifies $T_2 - T_1 \leq \Delta t$ where $T_2$ is the message receiving time. If time delay in message transmission is invalid, $S$ denies the request. Otherwise, $S$ computes $X^d~\mbox{mod}~N = (CID||k||ID_U)$, $H = h(d\oplus ID_U)$ and $R = h(h(H\oplus k), k, ID_U, T_1)$.

\item [\bf Step 2.] Server $S$ checks $CID \overset{?}{=}~h(H\oplus k)$ and $R \overset{?}{=}~R'$. If the verification does not hold, $S$ terminates the session. Otherwise, $S$ computes $\lambda = h(H, CID, R, T_1, T_2)$ and $V = h(\lambda, H, T_1, T_2)$ then responds with message $<V, T_2>$ to $U$.

\item[\bf Step 3.] Upon receiving the message $<V, T_2>$, $SC$ verifies $T_3 - T_4 \leq \Delta t$. If the verification fails, $SC$ terminates the session. Otherwise, $SC$ computes $\lambda' = h(H, CID, R, T_1, T_2)$ and $V' = h(\lambda', H, T_1, T_2)$. Then $SC$ verifies $V \overset{?}{=}~V'$. If condition holds, the verification is succeeded. Moreover, $U$ and $S$ consider $\lambda$ as the session key.

\end{description}
\subsection{Password Change Phase}

Any legal user $U$ can change the password by adopting the following steps.
\begin{description}

\item [\bf Step 1.] The user $U$ inputs his current password $PW_U$ and new password
$PW_{new}$.
\item [\bf Step 2.] The smart card $SC$ computes $v_{new} = v\oplus h(PW_U\oplus t)\oplus(PW_{new}\oplus t)$ and replace $v$ with $v_{new}$.

\end{description}

\section{Analysis of Lin's Scheme}\label{lin-crypt}

This section demonstrates the inefficiency of Lin's scheme login and password change phase and their consequences.

\renewcommand{\labelitemi}{$\bullet$}
\subsection{Inefficient password change phase}
If user inputs wrong password $PW_U^*$ instead of $PW_U$, then the smart card executes the password change phase as follows:

  \begin{itemize}
    \item  Compute $W^* = h(PW_U^*\oplus t)$ and then $v^* = v\oplus W^*\oplus h(PW_{new}\oplus t) = h(PW_U\oplus t)\oplus h(d\oplus ID_U)\oplus h(PW_U^*\oplus t)\oplus h(PW_{new}\oplus t)$.

    \item  Replace $v$ with $v^*$, where $v^* \neq h(d\oplus ID_U)\oplus h(PW_{new}\oplus t)$, as $PW_U^* \neq PW_U$.

  \end{itemize}

The smart card inability of wrong password detection, may cause \textbf{denial of service attack}, which is clear from the following facts:

 \begin{itemize}

 \item $U$ inputs $ID_U$ and $PW_{new}$ into smart card. Then, smart card chooses $k\in Z_N$ and computes $W = h(PW_{new}\oplus t)$, then $H^*$ is computed as:

   \begin{eqnarray*}
     H^* &=& v^*\oplus W \\
         &=& h(PW_U\oplus t)\oplus h(d\oplus ID_U)\oplus h(PW_U^*\oplus t)\oplus h(PW_{new}\oplus t)\oplus h(PW_{new}\oplus t) \\
      &=& h(PW_U\oplus t)\oplus h(d\oplus ID_U)\oplus h(PW_U^*\oplus t)\neq h(PW_U\oplus t)
   \end{eqnarray*}

 \indent It also computes $CID^* = h(H^*\oplus k)$, $R^* = h(CID^*, k, ID_U, T_1)$, and\\ $X^* = (CID^*|| k||ID_U)^e~\mbox{mod}~N$, where $T_1$ is the current time stamp. Finally, it sends the login message $<X^*, R^*, T_1>$ to $S$.

 \item Upon receiving $U$'s message at time $T_2$, $S$ verifies $T_2 - T_1 \leq \Delta t$. The validation holds, as the message is originally sent by user. Then, $S$ computes $X^d~\mbox{mod}~N = (CID^*||k||ID_U)$, $H = h(d\oplus ID_U)$, $R = h(h(H\oplus k), k, ID_U, T_1)$.

\item  $S$ verifies $CID^* =? ~h(H\oplus k)$ and $R =? R^*$. The verification fails, since $CID^* = h(h(PW_U\oplus t)\oplus h(d\oplus ID_U)\oplus h(PW_U^*\oplus t)\oplus k)$ as $PW_U^* \neq PW_U$. Moreover, $R \neq R^*$ as $H^* \neq h(d\oplus ID_U)$.

\item $S$ terminates the session as message authentication fails.

\end{itemize}
In this case, user can never login into the server because of denial of service scenario for legal user. 



\renewcommand{\labelitemi}{$-$}
\subsection{Inefficient login phase:}
An efficient smart card based remote user authentication scheme should be able to identify incorrect login request so that extra communication and computation cost should not occur due to incorrect login.  However, in Lin's scheme, the smart card does not verify the correctness of input. And, if a patient inputs incorrect identity or password, the smart card executes the login session. In this case, the following cases are possible:


\noindent{\bf Case 1.}
 If user inputs correct identity  $ID_U$ and incorrect password $PW_U^*$. Then without verifying the input, the smart card executes the login phase as follows:

\begin{itemize}
 \item The smart card chooses $k\in Z_N$ and computes $W^* = h(PW_U^*\oplus t)$, then $H^*$ as:

   \begin{eqnarray*}
     H^* &=& v\oplus W^* \\
         &=& h(PW_U\oplus t)\oplus h(d\oplus ID_U)\oplus h(PW_U^*\oplus t)
   \end{eqnarray*}

  $U$ also computes $CID^* = h(H^*\oplus k)$, $R^* = h(CID^*, k, ID_U, T_1)$ and\\ $X^* = (CID^*||k||ID_U)^e~\mbox{mod}~N$, where $T_1$ is the current time stamp. Finally, it sends the login message $<X^*, R^*, T_1>$ to $S$.

 \item Upon receiving the message, $S$ verifies $T_2 - T_1 \leq \Delta t$, where $T_2$ is $S$'s message receiving time. If time delay in message transmission is valid, $S$ computes $(X^*)^d~\mbox{mod}~N = (CID^*||k||ID_U)$, $H = h(d\oplus ID_U)$, $R = h(h(H\oplus k), k, ID_U, T_1)$.

\item  $S$ verifies $CID^* =? h(H\oplus k)$ and $R =? R^*$. The verification fails, since $H^* = h(PW_U\oplus t)\oplus h(d\oplus ID_U)\oplus h(PW_U^*\oplus t)$ instead of $h(d\oplus ID_U)$ as $PW_U^* \neq PW_U$. Moreover, $R \neq R^*$, as $H^* \neq h(d\oplus ID_U)$.

\item $S$ terminates the session, as message authentication does not hold.

\end{itemize}

\noindent{\bf Case 2.} If a user inputs incorrect identity $ID_U$ and correct password $PW_U$. However, the smart card does not verify the correctness of input identity and executes the session, which works as follows:
\begin{itemize}
\item Smart card chooses $k\in Z_N$. It computes $W = h(PW_U\oplus t)$, $H = v\oplus W$, $CID = h(H\oplus k)$, $R^* = h(CID, k, ID_U^*, T_1)$ and $X^* = (CID||k||ID_U^*)^e~\mbox{mod}~N$, where $T_1$ is the current time stamp. Finally, it sends login message $<X^*, R^*, T_1>$ to $S$.

\item Upon receiving the message $<X^*, R^*, T_1>$, $S$ verifies $T_2 - T_1 \leq \Delta t$, where $T_2$ is the message receiving time. If time delay in message transmission is valid, $S$ computes $X^d\mbox{mod}~N = (CID||k||ID_U^*)$, $H' = h(d\oplus ID_U^*)$, $R' = h(h(H'\oplus k), k, ID_U^*, T_1)$.

\item When, $S$ verifies $CID =? h(H'\oplus k)$, the verification does not hold, since $H = h(d\oplus ID_U)$ while $H' = h(d\oplus ID_U^*)$ where $ID_U^* \neq ID_U$. Moreover, the condition $R =? R'$ is not satisfied.

\item Finally, $S$ terminates the session.
\end{itemize}
\noindent{\bf Case 3.}
If User inputs incorrect identity and password, then also the smart card executes the login session as follows:

\begin{itemize}
 \item The smart card chooses $k\in Z_N$ and computes $W^* = h(PW_U^*\oplus t)$, then $H^*$ as:

   \begin{eqnarray*}
     H^* &=& v\oplus W^* \\
         &=& h(PW_U\oplus t)\oplus h(d\oplus ID_U)\oplus h(PW_U^*\oplus t)\\
      &=& h(PW_U\oplus t)\oplus h(d\oplus ID_U)\oplus h(PW_U^*\oplus t)
   \end{eqnarray*}

  $U$ also computes $CID^* = h(H^*\oplus k)$, $R^* = h(CID^*, k, ID_U^*, T_1)$ and\\ $X^* = (CID^*||k||ID_U^*)^e~\mbox{mod}~N$, where $T_1$ is the current time stamp. Finally, $U$ sends the login message $<X^*, R^*, T_1>$ to $S$.

 \item Upon receiving the message, $S$ verifies $T_2 - T_1 \leq \Delta t$, where $T_2$ is the message receiving time. If time delay in message transmission is valid, $S$ computes $(X^*)^d~\mbox{mod}~N = (CID^*||k||ID_U^*)$, $H' = h(d\oplus ID_U^*)$, $R' = h(h(H'\oplus k), k, ID_U^*, T_1)$.

\item  $S$ verifies $CID^* =? h(H'\oplus k)$, the verification fails because $H^* =  h(PW_U\oplus t)\oplus h(d\oplus ID_U)\oplus h(PW_U^*\oplus t)$ while $H' = h(d\oplus ID_U^*)$. Moreover, $R =? R^*$ also does not hold. So, the verification fails.

\item $S$ terminates the session, as message verification does not hold.

\end{itemize}

\noindent{\bf Case 4.} An attacker can also intercept and modify the login message, which is justified as follows:

\begin{itemize}
  \item An attacker intercepts $U$'s login message  $<X, R, T_1>$.
  \item Replace the message with $<X, R^*, T_1>$, where $R^* = h(R||T_1)$.
  \item On receiving the message, when $S$ verifies $R =? R^*$. The verification fails because $R = h(h(H\oplus k), k, ID_U, T_1)$ while $R^* = h(R||T_1)$. Thus, $S$ terminates the session.
\end{itemize}

 In both the cases, wrong password input and message impersonation attack, session terminates. And, user may not identify the correct reason of session failure.



\section{Review of Cao and Zhai's Scheme}\label{rc}

In 2013, Cao et al.~\cite{cao2013improved}  proposed an improvement of Chen el al.'s dynamic ID-based authentication scheme for telecare medical information systems~\cite{chen2012efficient}. In their scheme, the server initially chooses two large prime numbers $p$ and $q$, where $p \equiv 3 \pmod 4$ and $q \equiv 3 \pmod 4$. The server keeps $p$ and $q$ secret and makes $n = pq$ public. The Cao and Zhai's Scheme comprises the following phases:

 \begin{itemize}
  \item (i)~ Registration.
  \item (ii)~Authentication and key agreement.
  \item (iii)~Password change
  \item (iv)~Lost smart card revocation
\end{itemize}

\subsection{Registration Phase}

\renewcommand{\labelitemi}{$\bullet $}
\begin{description}

\item [\bf Step 1.] $U$ selects  an appropriate identity $ID_U$, password $PW_U$ and chooses a random number $b$. Then, he calculates $W = h(b||PW_U)$ and submits $\{ID_U, W\}$ to $S$.

\item[\bf Step 2.] $S$ verifies $ID_U$ and the user account database. If the patient is unregistered, $S$ stores $ID_U$ and $N = 0$ into the $U$'s account database. 

 \item [\bf Step 3.] $S$ calculates $J = h(p||q||ID_U||N)$ and $L = J \oplus W$. Finally, it embeds the parameters $(L, n, h(\cdot))$ into smart card and issues the smart card to $U$ .

    \item [\bf Step 4.] Upon receiving the smart card, $U$ stores $b$ into the smart card.

\end{description}

\subsection{Authentication and key agreement phase}
$U$ inserts his smart card into the card reader and inputs his identity and password to execute the login.

\begin{description}

\item [\bf Step 1.] $SC$ chooses a random number $r_u$ and calculates $J = L\oplus h(b||PW_U)$ and $AID = (ID_U ||J ||r_u)^2 \pmod n$. Then, it sends the login message $\{AID\}$ to $S$.

 \item[\bf Step 2.] Upon receiving the login message, $S$ decrypts $\{AID\}$ using its private keys $p$ and $q$ and get $(ID_U^{'}||J^{'}||r^{'})$. $S$ verifies the validity of $ID_U^{'}$. If verification holds, $S$ extracts the corresponding value $N$ from its database. Then, it achieves $J =h(p||q||ID_U^{'}||N)$ and verifies $J =?~J^{'}$. If verification holds, $S$ chooses a random number $r_s$ and computes $K_s = h(r_u ||r_s)$ and $C_s = h(K_s ||r_s )$. Finally, it sends $\{r_s, C_s\}$ to $U$.

\item[\bf Step 3.] Upon receiving $\{r_s, C_s\}$, $U$ achieves $K_u = h(r_u ||r_s)$ and verifies $C_s =?~h(K_u ||r_s)$. If condition does not hold, $U$ terminates the session. On the contrary, $U$ identifies $K_u$ as the session key and sends $C_u = h(r_s ||K_u)$ to $S$.

\item[\bf Step 4.] Upon receiving $C_u$, $S$ verifies $C_u =?~h(r_s ||K_s)$. If condition does not hold, $S$ stops the session. On the contrary, $S$ decides $K_s$ as the session key.
\end{description}

\subsection{Password Change Phase}

A legal user $U$ can change the password by inputting the identity $ID_U$, the old password $PW_U$ and the new password $PW_{new}$ as follows.
\begin{description}

\item [\bf Step 1] This step follows {\em Step 1} \& {\em 2} of {\em Authentication and key agreement phase}.
\item [\bf Step 2. ] Upon achieving $\{r_s, C_s\}, U$ computes $K_u = h(r_u||r_s)$ and verifies $C_s =?~h(K_u ||r_s)$. If condition does not hold, $SC$ terminates the session. Otherwise, it computes $L_{new} = L\oplus h(b||PW_U)\oplus h(b||PW_{new})$ and updates $L$ with $L_{new}$.

\end{description}
\subsection{Lost Smart Card Revocation}

If a legal user lost his smart card, he  can achieve a new smart card from server as follows:
\begin{description}

\item [\bf Step 1. ] $U$ selects a password $PW_U$ and a random number $b$. $U$ computes $W = h(b||PW_U)$ and sends $\{ID_U, W\}$ to $S$.
\item [\bf Step 2. ] Let $N = N + 1$, then $S$ achieves $J = h(p||q||ID_U||N)$ and $L =J \oplus W$. Finally, $S$ embeds $(L, n, h(\cdot))$ into the smart card and issues the card to $U$ while $U$ stores $b$ into the smart card.

\end{description}


\renewcommand{\labelitemi}{$-$}
\section{Analysis of Cao and Zhai's Scheme}\label{cc}
In this section, we will show that Cao and Zhai's~\cite{cao2013improved} scheme does not provide efficient login and user-friendly password change phase. Additionally, the scheme is vulnerable to known session specific temporary information attack and replay attack.

\subsection{Inefficient login phase:}
A smart card based authentication scheme should be able to detect incorrect login so that unnecessary computation can be avoided~\cite{das2013secure}. However, Cao and Zhai's scheme, smart card do not verify the correctness of input parameters (identity and password). A user can input wrong input, as human may forget or use one account identity into another account or by mistake he may enter a wrong input. In case of incorrect login (wrong identity and password input), smart card executes the session in Cao and Zhai's scheme. Then, the following cases take place:

\noindent{\bf Case 1.} If a user inputs the correct identity and wrong password, then login session executes as follows:

\begin{itemize}

 \item $U$ inputs correct identity $ID_U$ and wrong password $PW_U^*$ into the smart card. Without verifying the correctness of input parameters, the smart card selects a random number $r_u$ and achieves $J^* = L\oplus h(b||PW_U^*) =  h(p||q||ID_U||N)\oplus h(b||PW_U)\oplus h(b||PW_U^*)$.

 \item Smart card also computes $AID^* = (ID_U||{J^*}||r_u)^2~\pmod n$, then sends the login message $<AID^*>$ to $S$.

 \item  Upon receiving the message $<AID^*>$, $S$ decrypts ${AID^*}$ and achieves $(ID_U||{J^*}||r_u)$. $S$ verifies the correctness of $ID_U$. The verification holds, as $ID_U$ is correctly entered. Then $S$ retrieves the corresponding value $N$. It computes $J = h(p||q||ID_U||N)$ and verifies $J =?~J^*$.

\item  The verification does not hold, since $J^*= h(p||q||ID_U||N)\oplus h(b||PW_U)\oplus h(b||PW_U^*)$. Then, $S$ terminates the session.

\end{itemize}

\noindent{\bf Case 2.} If the user enters incorrect identity and correct password. In this case, there are two possibilities:
 \begin{enumerate}
   \item Input incorrect identity may match with some other legitimate user identity, as a mistaken identity can be the legitimate identity of any valid user.
   \item Input incorrect identity may not match with any user identity.
 \end{enumerate}

\noindent{\bf Case 2.1.}  When user $U$ inputs incorrect identity $ID_U^*$ and correct password $PW_U$ into the smart card. If incorrect identity matches with some legitimate user identity in the system the process executes as follows:
\begin{itemize}

 \item Without verifying the correctness of input parameters, the smart card selects a random number $r_u$ and achieves $J = L\oplus h(b||PW_U) =  h(p||q||ID_U||N)\oplus h(b||PW_U)\oplus h(b||PW_U) = h(p||q||ID_U||N)$.

 \item Smart card also computes $AID^* = (ID_U^*||{J}||r_u)^2~\pmod n$, then sends the login message $<AID^*>$ to $S$.

 \item  Upon receiving the message $<AID^*>$, $S$ decrypts ${AID^*}$ and achieves $(ID_U^*||{J}||r_u)$. $S$ verifies the correctness of $ID_U$. If $ID_U^*$ matches with some legitimate user identity, the verification holds. Then, $S$ retrieves the corresponding value $N^*$ and computes $J^* = h(p||q||ID_U^*||N^*)$ and verifies $J =?~J^*$. The verification does not hold, as $ID_U\neq ID_U^*$. So, $S$ terminates the session.

\end{itemize}

\noindent{\bf Case 2.2.} If input incorrect identity does not match with any legitimate user identity.
\begin{itemize}

 \item $U$ inputs incorrect identity $ID_U^*$ into the smart card. Then, the smart card selects a random number $r_u$ and achieves $J = L\oplus h(b||PW_U) =  h(p||q||ID_U||N)\oplus h(b||PW_U)\oplus h(b||PW_U) = h(p||q||ID_U||N)$.

 \item Smart card also computes $AID^* = (ID_U^*||{J}||r_u)^2~\pmod n$, then sends the login message $<AID^*>$ to $S$.

 \item  Upon receiving the message $<AID^*>$, $S$ decrypts ${AID^*}$ and achieves $(ID_U^*||{J}||r_u)$. $S$ verifies the correctness of $ID_U^*$. The verification does not hold, as the identity does not match with any legitimate user identity.
\end{itemize}

\noindent{\bf Case 3.} If both input parameters (identity and password) are incorrect, then {\em Case 1} or {\em Case 2.2} will repeat.

%
%


\renewcommand{\labelitemi}{$\bullet$}
\subsection{Unfriendly password change phase}
A user should be allowed to change the password without server assistance~\cite{li2013enhanced}. However, in Cao and Zhai's scheme, a user cannot change the password independently without server's assistance. To change the password, the user has to communicate with the server each time. This makes the password change phase unfriendly.


\section{Review of Xie et al.'s Scheme}\label{rx}

In 2013, Xie et al.~\cite{lin2013security}  proposed an improvement of Chen el al.'s dynamic ID-based authentication scheme for telecare medical information systems~\cite{chen2012efficient}. Their scheme is based on RSA public-key cryptosystem~\cite{rivest1978method}. In which, the medical server S chooses two large
primes $p$ and $q$, then computes $n = pq$. It also chooses an integer $e$, which is relatively prime to $(p-1)(q-1)$ and derives $X$ such that $eX \equiv 1 ~\pmod(p-1)(q-1)$.   Xie et al.'s scheme also has the similar phases as Cao and Zhai's scheme. Xie et al.'s Scheme comprises the following phases:

\renewcommand{\labelitemi}{$ $}
 \begin{itemize}
  \item (i)~ Registration.
  \item (ii)~Login.
  \item (iii)~Verification.
  \item (iv)~Password change.
  \item (iv)~Smart card revocation.

\end{itemize}


\subsection{Registration Phase}

A patient can registers his identity with the server and gets the smart card as follows:\\
\renewcommand{\labelitemi}{$\bullet $}
\begin{description}

\item [\bf Step 1.] $U$ selects his identity $ID_U$ and password $PW_U$, then submits $ID_U$ with registration request to the medical server.

\item[\bf Step 2.] Upon receiving the $U$'s message, $S$ checks the legitimacy of $ID_U$. If this is invalid, it terminates the session. Otherwise, it assigns a card number $SC$, then computes $J =  h(X||ID_U||N||SC)$. It also keeps the account table, which includes $(ID_U, RID) = (ID_U, E_X(ID_U,$\\
     $N, SC))$ to maintain the record of patient registration identity $ID_U$, card number $SC$, registration time $N$ (If patient is newly registered N=0, otherwise, N = N+1).

 \item [\bf Step 3.] $S$ embeds the values $\{ID_U, SC, N, J, n, e, h(\cdot)\}$ into the smart card and then returns the medical smart card to $U$ via secure channel.

    \item [\bf Step 3.] $U$ computes $L = J\oplus h(PW_U)$ and then replace  $J$ with $L$. 

\end{description}

\subsection{Login Phase}
When a patient wishes to login into TMIS server, he inserts his smart card into the card reader, then login session executes as follows:

\begin{description}

\item [\bf Step 1.] The smart card generates a random number $a$ and achieves $J = L\oplus h(PW_U), A= J^a \pmod n$, $C_1 = h(T_u||J||A)$ and $AID = (ID_U||N||SC||C_1||A)^e~\pmod n$, where $T_u$ is the current timestamp.

 \item[\bf Step 2.] Smart card sends the login message $<AID, T_u>$ to $S$.

\end{description}

\subsection{Verification Phase}
Smart card and server performs the following steps to authenticate each other:

\begin{description}

\item [\bf Step 1. ] Upon receiving the message $<AID, T_u>$ at time $T$, $S$ Verify $T - T_u \leq \Delta t$, where $\Delta t$ is the valid time delay in message transmission. If time delay is invalid, terminate the session. Otherwise, $S$ computes $AID^X\pmod n = (ID_U||N||SC||C_1||A)$ and then $J = h(X||ID_U||N||SC)$ and $C_1' = h(T_u||J||A)$. It also decrypts $RID$ and achieves $(ID_U, N, SC)$.

\item [\bf Step 2. ] $S$ verifies the validity of $ID_U$, $SC$ and $C_1'$. If verification does not hold, terminate the session. Otherwise, $S$ generates a random number $b$ and then computes $B = J^b~\pmod n$, $C = A^b ~\pmod n = J^{ab}~\pmod n$ and $C_2 = h(C_1||C||T_s||B)$. Finally, $S$ sends the message $\{C_2, T_s, B\}$ to $U$, where $T_s$ is the current timestamp.

\item[\bf Step 3.] Upon receiving $S$'s message $\{C_2, T_s, B\}$ at time $T'$, smart card verifies $T' - T_s \leq \Delta t$. If verification fails, terminate the session. Otherwise, it computes $C = B^a ~\pmod n = J^{ab}~\pmod n$ and $C_2' = h(C_1||C||T_s||B)$. Then , it verifies $C_2 =? C_2'$. If it holds, $U$ computes $sk = h(T_u||C||T_s)$.

\end{description}


\subsection{Password Change Phase}

Any legal user $U$ can change the password by using the
following steps.
\begin{description}

\item [\bf Step 1. ] $U$ inputs his old password $PW_U$ and new password $PW_{new}$ into his smart card.
\item [\bf Step 2. ] Smart card computes $L_{new} = L\oplus h(PW_U)\oplus h(PW_{new})$, then replaces $L$ with $L_{new}$.

\end{description}
\subsection{Smart Card Revocation}

If a legal user lost his smart card, then user can get a new smart card from server as follows:
\begin{description}

\item [\bf Step 1. ] $U$ submits its new smart card request with $ID_U$ to $S$.
\item [\bf Step 2. ] $S$ verifies the validity of $U$'s  ID card. Then, server adopts the same procedure as in registration phase. Additionally, it replaces $N$ with $N+1$ and then $RID = E_X(ID_U, N, SC)$ with $RID' = E_X(ID_U, N, SC')$ in account table, where $SC'$ is a new smart card number.

\end{description}


\section{Analysis of Xie et al.'s Scheme}\label{cx}

In this section, we will show that Xie et al.'s scheme  does not provide efficient login and password change phase.

\renewcommand{\labelitemi}{$-$}
\subsection{Inefficient login phase:}
In Xie et al.'s scheme, smart card does not verify the correctness of input. However, a human may sometimes input wrong password due to mistake.
If user inputs incorrect password even then the smart card executes the login phase, which is as follows:

\begin{itemize}

 \item $U$ inputs incorrect password $PW_U^*$ into smart card. Then, the smart card generates a random number $a$ and achieves $J^* = L\oplus h(PW_U^*) =  h(X||ID_U||N||SC)\oplus h(PW_U)\oplus h(PW_U^*)$.

 \item Smart card also computes $A^* = {J^*}^a~\pmod n$, $C^*_1 = h(T_u||J^*||A^*)$ and $AID^* = (ID_U||N||SC||$\\
     $C^*_1||A^*)^e~\pmod n$, where $T_u$ is the current time stamp.

 \item Finally, smart card sends the login message $<AID^*, T_u>$ to $S$.

 \item  Upon receiving the message $<AID^*, T_u>$ at time $T$, $S$ verifies $T - T_u \leq \Delta t$. The verification succeeds, as $T_u$ is the valid timestamp. Then, $S$ computes ${AID^*}^X~\pmod n = (ID_U||N||SC||C^*_1||A^*)$ and then $J = h(X||ID_U||N||SC)$ and $C_1 = h(T_u||J||A^*)$. It also decrypts $RID$ and achieves $(ID_U, N, SC)$.

\item  $S$ verifies the validity of $ID_U$, $SC$ and $C_1 =? C^*_1$. The verification does not hold, as  $C_1 \neq C^*_1$ since $J\neq J^*$.

\item $S$ terminates the session as message authentication does not hold.

\end{itemize}

An attacker can track, intercept, and record the messages. Let the attacker has achieved an previously transmitted message $<AID, T_u>$.
An attacker can also modify and replace the messages that transmit via public channel.

In case if user inputs correct password. Although, the attacker intercepts the message $<AID, T_u>$ and replaces with old transmitted message $<AID', T_E>$ as an attacker can intercept, modify,record, and replace the message that transmit via public channel, where $AID'$ is the old transmitted value and $T_E$ is current timestamp. Then, on receiving the message $<AID', T_E>$, server performs the following steps:

\begin{itemize}

\item S verifies $T - T_E \leq \Delta t$, where $T$ is the message receiving time. The verification succeeds, as attacker usage its current timestamp.

\item S computes ${AID'}^X~\pmod n = (ID_U||N||SC||C'_1||A')$, where $A' = {J}^a~\pmod n$ and $C_1 = h(T'_u||J||A')$  for old timestamp $T'_u$. Then, compute $J = h(X||ID_U||N||SC)$ and $C_1 = h(T_E||J||A)$. It also decrypts $RID$ and achieves $(ID_U, N, SC)$.

\item  $S$ verifies the validity of $ID_U$, $SC$ and $C_1 =?~C'_1$. The verification of $C_1 =?~C'_1$ does not hold, as $T'_E \neq T_u$.

\item Finally, $S$ terminates the session as message authentication does not hold.

\end{itemize}

In \textbf{case 1},  in spite of wrong password the smart card executes the login session. However, server terminates the session. Moreover, in case of message impersonation attack (\textbf{case 2}), the server also terminates the session in same step. Therefore, if the session terminates, it will be difficult for user to identify that exactly why the session has terminated.
%
%
%
%
%


\renewcommand{\labelitemi}{$\bullet$}
\subsection{Inefficient password change phase}
  User can change his password independently without server's assistance. However, the smart card does not verify the user's password. When a user inputs the wrong password, the smart card does not verify the correctness of password and processes the password change request. However, a user may enter a wrong password, as human may sometimes forget the password or commit some mistake. This may cause the denial of service scenario where a user will no longer communicate with server using the same smart card.

  If user inputs wrong password $PW_U^*$ instead of $PW_U$, then the smart card executes the password change phase as follows:

  \begin{itemize}

    \item  Compute $h(PW_U^*)$ and then achieve $J^*= L\oplus h(PW_U^*) = J\oplus h(PW_U)\oplus h(PW_U^*)$, as $PW_U \neq PW_U^*$.

   \item  Compute $L_{new} = J^*\oplus h(PW_{new}) = J\oplus h(PW_U)\oplus h(PW_U^*)\oplus h(PW_{new})$  and then replace $L$ with $L_{new}$.

  \end{itemize}

The smart card inability of wrong password detection, may cause \textbf{denial of service attack}, which is clear from the following facts:

 \begin{itemize}

 \item $U$ inputs $PW_{new}$ into smart card. Then, the smart card generates a random number $a$ and computes $J' = L_{new}\oplus h(PW_{new}) =  J\oplus h(PW_U)\oplus(PW_U^*)$ $\neq J = h(X||ID_U||N||SC)$.

 \item Smart card also computes $A' = {J'}^a~\pmod n$, $C'_1 = h(T_u||J'||A')$ and $AID' = (ID_U||N||SC||C'_1$\\
 $||A')^e~\pmod n$ with current timestamp $T_u$. Then, it sends the login message $<AID', T_u>$ to $S$.

 \item  Upon receiving the message $<AID', T_u>$ at time $T$, $S$ verifies $T - T_u \leq \Delta t$, which holds, as $T_u$ is the valid timestamp. Then, $S$ computes ${AID}^X \pmod n = (ID_U||N||SC||C^*_1||A^*)$ and then $J = h(X||ID_U||N||SC)$ and $C_1 = h(T_u||J||A')$. It also decrypts $RID$ and achieves $(ID_U, N, SC)$.

\item  $S$ verifies the validity of $ID_U$, $SC$ and $C_1 =? C^*_1$. The verification of $C_1 =? C^*_1$ does not hold, as  $J\neq J'$. Then, it terminates the session. 

\end{itemize}

The above discussion indicates that user can never login to the server after inputting the wrong password in password change phase. This causes \textbf{denial of service attack} for legal user. Moreover, user can not track the mistake, as the message also terminates because of inefficient login phase.


\section{Review of Xu et al.'s Authentication and Key Agreement Scheme Based on ECC for Telecare Medicine Information Systems}\label{xu-review}

In 2014, Xu et al.~\cite{xu2014secure} presented a two-factor authenticated key agreement  Scheme based on ECC for TMIS. Their scheme comprises four phases, namely, registration phase, login phase, verification phase and password update phase. 

Before initiating the registration phase, the server $S$ opts an elliptic curve $E: y^2 =x^3 + ax + b (mod~p)$ and generates a $n$th order group $E_p(a, b)$  over the prime field $Z_p^*$, where n is a large prime number. $S$ chooses a base point $P =(x_0, y_0)$ such that $n{\cdot}P = O$. $S$ selects a random number $s\in Z_p^*$ and compute $Y = s{\cdot}P$. S also chooses two one-way hash functions $h(\cdot): \{0, 1\}^*\rightarrow Z_p^*$ and $h_1(\cdots): G_p\times G_p \rightarrow Z_p^*$, where $ G_p$ is a additive elliptic curve group.  $S$ keeps $s$ secret.


\subsection{Registration Phase}

A user $U$ registers his identity $ID_U$ to the medical server $S$ and gets the medical smart card $SC$ with personalized security parameters as follows:
\begin{description}

\item [\bf Step 1.] User $U$ selects a password  $PW_U$ and a random number $r$. $U$ computes $W = h(PW_U||r)$ and sends the message $<ID_U, W>$ to $S$ via secure channel.

\item[\bf Step 2.] Upon receiving $U$'s request, $S$ computes $H = h(s\oplus ID_U)$ and $B = H\oplus W$. $S$ personalizes $U$'s smart card by embedding the parameters $\{E_p, P, Y, B, h(\cdot), h_1(\cdot)\}$, and sends the smart card to $U$ via secure channel.

    \item [\bf Step 3.] Upon receiving the smart card, $U$ stores $r$ into the smart card.

\end{description}

\subsection{Login Phase}
  The user $U$ inserts his smart card $SC$ into the card reader, and inputs $ID_U$ and $PW_U$. $SC$ computes $W = h(PW_U||r)$, $H = B\oplus W$, $C_1 = a{\cdot}P$, $C_2 = a{\cdot}Y$, $CID = ID_U\oplus h_1(C_2)$ and $F = h(ID_U||H||T_1)$ for secret nonce $a$ selected by $U$ and  $T_1$ is the current timestamp. Finally, $SC$ sends the login message $<C_1, CID, F, T_1>$  to $S$.

\subsection{Verification Phase}
User $U$ and server $S$ verify the authenticate each other and compute session key as follows:


\begin{description}

\item [\bf Step 1.] Upon receiving the message $<C_1, CID, F, T_1>$, $S$ verifies $T_1$. If $T_1$ is a null timestamp, $S$ terminates the session. Otherwise, $S$ computes $C_2' = s{\cdot}C_1$, $ID_U' = CID\oplus h_1(C_2)$ and $H' = h(s\oplus ID_U')$. Then $S$ verifies $F \overset{?}{=} h(ID_U'||H'||T_1)$. If verification does not hold, the session is terminated. Otherwise, $S$ considers $U$ as a legitimate user.

\item [\bf Step 2.] Server $S$ chooses a nonce $C$, and computes $D_1 = c{\cdot}P$, $D_2 = c{\cdot}C_1$, $sk = h(ID_U'||h_1(D_2)||H')$ and $G = h(sk||H'||T_2)$ and then responds with message $<D_1, G, T_2>$ to $U$, where $T_2$ is the current timestamp.

\item[\bf Step 3.] Upon receiving the message $<D_1, G, T_2>$, $SC$ verifies the validity of $T_2$. If the verification holds, $SC$ computes $D_2' = a{\cdot}D_1$ and $sk' = h(ID_U||h_1(D_2')||H)$. Then $SC$ verifies $G \overset{?}{=}~h(sk'||H||T_2)$. If  the verification succeeds, $U$ considers S as an authorized server. Moreover, $U$ and $S$ consider $sk$ as the session key.

\end{description}
\subsection{Password Change Phase}

Any user $U$ can change the password by adopting the following steps.
\begin{description}

\item [\bf Step 1.]  $U$ inputs $PW_U$ and $ID_U$. Then $SC$ computes $W = h(PW||r)$  and $H = B\oplus W$.
\item [\bf Step 2.] $U$ is asked to enter a new password $PW_{new}$. Upon receiving $PW_{new}$, $SC$ computes $W_{new} = h(PW_{new}||r)$ and $B_{new} = H\oplus W_{new}$, and replaces $B$ with $B_{new}$.

\end{description}

\section{Flaws in Xu et al.'s Scheme}\label{xu-crypt}

This section demonstrates the inefficiency of login and password change phases of Xu et al.'s scheme and their disadvantages.

\subsection{Inefficient password change phase}
 Xu et al.'s scheme executes the password change phase without verify the correctness of input. If a user commits some mistake in entering current password, outcome will be denial of service attack.  The following discussion justified the claim:

  \begin{itemize} 

  \item A user $U$ inputs identity $ID_U$ and incorrect current password $PW_U^*$.

    \item  The smart card computes $W^* = h(PW_U^*||r)$ and then $H^* = B\oplus W^* = H\oplus h(PW_U||r)\oplus h(PW_U^*||r) \neq H$ as $PW_U^* \neq PW_U$.

    \item $U$ inputs a new password  $PW_{new}$ and smart card computes $W_{new} = h(PW_{new}||r)$ and $B_{new}^* = H^*\oplus W_{new}$.

  \end{itemize}

The inability of incorrect input detection, leads to denial of service scenario for an authorized user as follows:

The user $U$ inserts his smart card $SC$ into the card reader, and inputs $ID_U$ and $PW_U$. $SC$ computes $W = h(PW_U||r)$, $H = B\oplus W$, $C_1 = a{\cdot}P$, $C_2 = a{\cdot}Y$, $CID = ID_U\oplus h_1(C_2)$ and $F = h(ID_U||H||T_1)$ for secret nonce $a$ and  current timestamp $T_1$. Finally, $SC$ sends the login message $<C_1, CID, F, T_1>$  to $S$.

 \begin{itemize}

 \item To login, $U$ inputs identity  $ID_U$ and new changed password $PW_{new}$. $SC$ computes $W_{new} = h(PW_{new}||r)$ and then $H^* = B_{new}^*\oplus W_{new}$.

\item $SC$ also computes $C_1 = a{\cdot}P$, $C_2 = a{\cdot}Y$, $CID = ID_U\oplus h_1(C_2)$ and $F^* = h(ID_U||H^*||T_1)$ for secret nonce $a$ and  current timestamp $T_1$. Finally, $SC$ sends the login message $<C_1, CID, F^*, T_1>$  to $S$.

\item Upon receiving the message $<C_1, CID, F^*, T_1>$, $S$ verifies $T_1$. The verification holds as $U$ usage current timestamp. Then $S$ computes $C_2' = s{\cdot}C_1$, $ID_U' = CID\oplus h_1(C_2)$ and $H' = h(s\oplus ID_U')$. Then $S$ verifies $F^* \overset{?}{=} h(ID_U'||H'||T_1)$. The verification does not hold as $H^* \neq H$. The session is terminated.

\end{itemize}
The above discussion shows that user cannot login to the medical server using the same smart card due to flaws in password change phase. This shows that inefficient password change phase can lead to {denial of service attack for a legitimate user}.



\renewcommand{\labelitemi}{$-$}
\subsection{Inefficient login phase:}
 Xu et al.'s scheme does not check the correctness of input before executing the login session. Therefore, the login session can be  executed in case of incorrect input. This causes extra communication and computation overhead which is justified as follows:

\noindent{\bf Case 1.} User $U$ inputs correct identity  $ID_U$, but incorrect password $PW_U^*$ due to mistake. 

\begin{itemize}
 \item $SC$ computes $W^* = h(PW_U^*||r)$ and $H^* = B\oplus W^* = H\oplus h(PW_U||r)\oplus h(PW_U^*||r) \neq H$ as $PW_U^* \neq PW_U$

 \item $SC$ also computes $C_1 = a{\cdot}P$, $C_2 = a{\cdot}Y$, $CID = ID_U\oplus h_1(C_2)$ and $F^* = h(ID_U||H^*||T_1)$ for secret nonce $a$ selected by $U$ and  $T_1$ is the current timestamp.  $SC$ sends the login message $<C_1, CID, F^*, T_1>$  to $S$.

 \item Upon receiving the message $<C_1, CID, F^*, T_1>$, $S$ verifies $T_1$. If $T_1$ verification hold, $S$ computes $C_2' = s{\cdot}C_1$, $ID_U' = CID\oplus h_1(C_2)$ and $H' = h(s\oplus ID_U')$. Then $S$ verifies $F^* \overset{?}{=} h(ID_U'||H'||T_1)$. The verification does not hold as $H^* \neq H'$. The session is terminated.

\end{itemize}

\noindent{\bf Case 2.} User $U$ inputs correct password $PW_U$ and incorrect identity $ID_U^*$ by mistake.
\begin{itemize}
\item $SC$ computes $W = h(PW_U||r)$, $H = B\oplus W$, $C_1 = a{\cdot}P$, $C_2 = a{\cdot}Y$, $CID = ID_U^*\oplus h_1(C_2)$ and $F'^* = h(ID_U^*||H||T_1)$ for secret nonce $a$ selected by $U$ and  $T_1$ is the current timestamp. Finally, $SC$ sends the login message $<C_1, CID^*, F^*, T_1>$  to $S$.

\item  Upon receiving the message $<C_1, CID, F'^*, T_1>$, $S$ verifies $T_1$. If $T_1$ verification hold, $S$ computes $C_2' = s{\cdot}C_1$, $ID_U^* = CID^*\oplus h_1(C_2)$ and $H'^* = h(s\oplus ID_U^*)$. Then $S$ verifies $F'^* = \overset{?}{=} h(ID_U^*||H'^*||T_1)$. The verification does not hold as $H'^* \neq H$. The session is terminated.

\end{itemize}



%


\section{Comparison} \label{comparision}

We compare some of the recently published password based schemes for TMIS~\cite{cao2013improved,chen2012efficient,lee2013secure,lin2013security,wei2012improved,zhu2012efficient,xie2013robust} in Table~\ref{t4}. If the scheme prevents attack or satisfies the property, the symbol {\sf '$\surd$'} is used and if it fails to prevent attack or does not satisfy the attribute, the symbol {\sf $\times$} is used.


\begin{table}[H]
{
  \caption{Security attributes comparison with some password based schemes for TMIS}\label{t4}

\begin{tabular}{|l|c|c|c|c|c|c|c|c|c|} \hline
 \backslashbox{Security attributes}{Schemes}
    &\cite{wei2012improved}       &  \cite{zhu2012efficient}  & \cite{lee2013secure}   &  \cite{chen2012efficient} &\cite{cao2013improved} &  \cite{xie2013robust}  &  \cite{lin2013security}  & \cite{xu2014secure} \\ \hline

 User anonymity     & $\times$       & $\times$      &   $\surd$     & $\surd$  &   $\surd$   &   $\surd$ &   $\surd$   &   $\surd$    \\

Insider Attack    &   $\surd$     &   $\surd$       &   $\surd$     & $\surd$  &   $\surd$   &   $\surd$ &   $\surd$ &   $\surd$   \\








Replay attack         &   $\surd$  &   $\surd$                     &   $\surd$            & $\surd$  &   $\times$   &   $\surd$ &   $\surd$      &   $\surd$             \\





Session key agreement    &   $\surd$   & $\times$             &   $\surd$           & $\surd$  &   $\surd$   &   $\surd$ &   $\surd$  &   $\surd$     \\

Session key verification            &   $\surd$                  & $-$             &   $\surd$          &  $\times$  &   $\surd$   &   $\times$  &    $\times$ &   $\surd$ \\


Efficient password change       & $\times$                  & $\times$            & $\times$        & $\surd$  &   $\surd$   &  $\times$   &   $\times$ &   $\times$ \\

User-friendly password change      & $\times$                  & $\times$            & $\times$        & $\surd$  & $\times$    &   $\surd$ &   $\surd$ &   $\surd$ \\
Efficient login          & $\times$                  & $\times$            & $\times$        & $\surd$  &  $\times$     &   $\times$   &  $\times$ &   $\times$  \\
\hline
\end{tabular}
}

 \end{table}



\section{Conclusion}\label{conclusion}
It is clear from the discussion that inefficient login phase increases the number attempts to login into the server, which  would flood up the network with fake requests. As a result, the server may busy in indulging unnecessary calculations, which may cause wastage of time and resources. The findings also print out the drawback of inefficient login and password change phase. The study of login and password phase of authentication schemes for TMIS shows that a scalable scheme should maintain efficient login and password change phase to avoid unnecessary trouble.



%

\end{document}